\documentclass[11pt]{article}
\usepackage{amsmath}
\usepackage{amssymb}
\usepackage{amsthm,amsxtra}
\usepackage{epsf}
\usepackage{eepic}
\newlength{\mytopmargin}
\newlength{\myleftmargin}
\newtheorem{cor}{Corollary}

\newtheorem{prop}{Proposition}
\newcommand{\zz}{\mathbb Z}
\newcommand{\qq}{\mathbb Q}

\begin{document}

\vspace{4cm}
\noindent
{\bf Singularity dominated strong fluctuations for some random matrix
averages} 

\vspace{5mm}
\noindent
P.J.~Forrester${}^\dagger$ and
J.P.~Keating${}^*$

\noindent
${}^\dagger$
Department of Mathematics and Statistics,
University of Melbourne, 
Victoria 3010, \\
Australia ;
${}^*$School of Mathematics, University of Bristol, Bristol, BS8 1TW, UK

\small
\begin{quote}
The circular and Jacobi ensembles of random matrices have their
eigenvalue support on the unit circle of the complex plane and the
interval $(0,1)$ of the real line respectively. 
The averaged value of the modulus of the corresponding characteristic
polynomial raised to the power $2 \mu$  diverges, for $2\mu \le -1$,
at points approaching the eigenvalue support. Using the theory of generalized
hypergeometric functions based on Jack polynomials, the functional
form of the leading asymptotic
behaviour is established rigorously. In the circular ensemble case this
confirms a conjecture of Berry and Keating.
\end{quote}

\section{Introduction}
Random matrices from the classical groups --- $N \times N$ unitary matrices
$U(N)$, $N \times N$ real orthogonal matrices $O(N)$, and $N \times N$ unitary
matrices with real quaternion elements embedded as $2N \times 2N$ complex
unitary matrices $Sp(N)$ --- play a special role in the application of
random matrix theory to number theory (see e.g.~the recent review \cite{KS03}).
Of particular interest in such applications are the random matrix averages
of the modulus of the characteristic polynomial raised to some power
$2 \mu$ say. Thus in \cite{KS00a} it was shown how knowledge of
\begin{equation}\label{1.1}
\Big \langle | \det(zI - U) |^{2 \mu} \Big \rangle_{U \in U(N)}
\end{equation}
for $|z| = 1$ (in this case (\ref{1.1}) is in fact independent of $z$) allows
the mean value of the $2 \mu$-th power of the modulus of the Riemann zeta
function on the critical line to be predicted. Knowledge of the analogue of
(\ref{1.1}) for the classical groups $O(N)$ and $Sp(N)$ allows for similar
predictions in the case of families of $L$-functions 
\cite{CF00,KS00b,CFKRS02}.

Our interest is in the asymptotic behaviour of (\ref{1.1}) and its analogues
as $|z|$ approaches unity. Consider in particular the generalization of
(\ref{1.1}) 
\begin{equation}\label{1.1a}
\Big \langle \prod_{l=1}^N | z - e^{i \theta_l}
|^{2 \mu} \Big \rangle_{{\rm C}\beta{\rm E}_N},
\end{equation}
where C$\beta$E${}_N$ (circular $\beta$-ensemble)
refers to the eigenvalue probability density function
proportional to
\begin{equation}
\prod_{1 \le j < k \le N} | e^{ i \theta_k} - e^{ i \theta_j} |^\beta,
\qquad - \pi \le \theta_l < \pi.
\end{equation}
The cases $\beta = 1$ and $\beta = 4$ are well known in random matrix theory
as the COE (circular orthogonal ensemble) and CSE (circular symplectic
ensemble) respectively, while the case $\beta = 2$ is just
(\ref{1.1}). For $2 \mu \le -1$ it has been argued by Berry
and Keating \cite{BK02} that (\ref{1.1a}) diverges as 
$|1 - |z||^{-\delta}$ where, provided $N \gg |\mu|$, $\delta$ as a function of
$\mu$ is the non-smooth function
\begin{equation}\label{2.1}
\delta = {\rm int} [ {(k-1) \over \beta} + 1] 
\Big ( k-1 + {\beta \over 2} - {\beta \over 2} {\rm int} 
[ {(k-1) \over \beta} + 1] \Big ), \qquad k = 2 |\mu|,
\end{equation}
with int$[\cdot]$ denoting the integer  part. It was also suggested that for
$k$ values such that $(k-1) / \beta$ is an integer, there is a 
logarithmic correction to the behaviour of (\ref{1.1a}) which then diverges
as $|1 - |z||^{-\delta} \log | 1 - |z||$. The intricate behaviour exhibited by
(\ref{2.1}) is due to a phenomenom termed singularity dominated strong
fluctuations. Thus for $2 \mu \le -1$ and $|z|$ close to unity, degeneracies
of the spectrum of varying order are being probed \cite{BK02}. A similar
mechanism has been identified in the analysis of twinkling starlight
\cite{Be77}, van Hove-type singularities \cite{Be82}, and the influence of
classical periodic orbit bifurcations on quantum energy level \cite{BKS00}
and wavefunction statistics \cite{KP01}.

The arguments presented in \cite{BK02} leading to (\ref{2.1}) were
heuristic. Subsequently Fyodorov and Keating \cite{FK03} gave a rigorous
derivation for the exponent (\ref{2.1}) in the case $\beta = 1$,
$2 \mu \in \zz_{< 0}$, for the Gaussian orthogonal ensemble analogue of
(\ref{1.1a}),
\begin{equation}\label{3.1}
\Big \langle | \det ( xI - X) |^{2 \mu} \Big \rangle_{X \in {\rm GOE}_N}.
\end{equation}
Thus they proved that for $N \gg \mu$ and $k=2|\mu| \in \zz_{> 0}$
\begin{equation}
\Big \langle | \det ( i \epsilon I
 - X) |^{2 \mu} \Big \rangle_{X \in {\rm GOE}_N}
\mathop{\sim}\limits_{\epsilon \to 0} C_{N,k} \epsilon^{-k(k-1)/2}
\log {1 \over \epsilon},
\end{equation}
(the explicit evaluation of $C_{N,k}$ was also given). In this paper we will 
give a rigorous derivation of the exponent (\ref{2.1}), together with the
logarithmic corrections when present, for the average (\ref{1.1a}) in the
case of general $\beta > 0$ and general $2\mu \le -1$. Our analysis relies on
identifying (\ref{1.1a}) as a special generalized hypergeometric function
based on Jack polynomials \cite{Ya92,Ka93}. We are then able to use known
asymptotic properties of the latter to deduce the sought asymptotic 
behaviour of (\ref{1.1a}).

Also studied will be the $x \to 1^+$ asymptotic behaviour of
\begin{equation}\label{4.1}
\Big \langle \prod_{l=1}^N | x - x_l|^{2 \mu} \Big \rangle_{{\rm J}\beta
{\rm E}_N}
\end{equation}
where  J$\beta$E${}_N$ (Jacobi $\beta$-ensemble)
refers to the eigenvalue probability density function
proportional to
\begin{equation}\label{4.2}
\prod_{l=1}^N x_l^a (1 - x_l)^b \prod_{1 \le j < k \le N}
| x_k - x_j |^\beta, \qquad 0 < x_l < 1.
\end{equation}
In the special case $\beta = 2$, (\ref{4.1}) includes the analogue of
(\ref{1.1}) in the cases of the classical groups $O(2N+1)$, $O(2N)$ and
$Sp(N)$. Again our analysis proceeds via identifying (\ref{4.1}) as a
generalized hypergeometric function based on Jack polynomials.

As a final issue, it is pointed out in \cite{HKO01} that for fixed $|z| \ne 1$
and $N \to \infty$, Szeg\"o's theorem \cite{Sz52} on the
asymptotic behaviour of Toeplitz determinants with smooth generating
functions implies (\ref{1.1}) is a simple Gaussian in $\mu$,
\begin{equation}\label{1.9}
\lim_{N \to \infty} \Big \langle | \det (z I - U) |^{2 \mu}
\Big \rangle_{U \in U(N)} = e^{- \mu^2 \log |1 - |z|^2|}, \qquad
|z| < 1.
\end{equation}
This gives for the exponent characterizing the divergence as $|z| \to 1^-$,
$\delta = \mu^2$ and so does not reproduce (\ref{2.1}). Therefore the limit
$N \to \infty$ with $|z| \ne 1$ fixed has the effect of smoothing out the
singularity dominated strong fluctuations. Note that in this limit the
point $z$ is a macroscopic distance away from the eigenvalue support,
as measured in units of the inter-eigenvalue spacing. A theorem of
Johansson \cite{Jo88, Jo98}
will be used to establish the analogue of (\ref{1.9}) for the
average (\ref{1.1a}).

\section{Generalized hypergeometric functions based on Jack polynomials}
\subsection{Definitions}
\setcounter{equation}{0}
The Gauss hypergeometric function ${}_2 F_1$ is, for $|x| < 1$, defined by
the series
\begin{equation}\label{4.3}
{}_2 F_1(a,b;c;x) := \sum_{k=0}^\infty {(a)_k (b)_k \over (c)_k k!} x^k.
\end{equation}
A multivariable generalization of (\ref{4.3}) can be defined in which the sum
over the non-negative integers $k$ is replaced by a sum over
partitions $\kappa := (\kappa_1,\dots, \kappa_N)$, $\kappa_i \ge \kappa_j$
$(i < j)$ and $\kappa_i \in \zz_{\ge 0}$; the function
$$
(u)_k := u(u+1) \cdots (u+k-1) = {\Gamma(u+k) \over \Gamma(u)}
$$
is replaced by
$$
[u]_\kappa^{(\alpha)} := \prod_{j=1}^N {\Gamma(u-(j-1)/\alpha + \kappa_j) \over
\Gamma(u - (j-1)/\alpha) };
$$
$k!$ is replaced by $|\kappa|! := (\sum_{j=1}^N \kappa_j)!$; and $x^k$ is
replaced by the homogeneous symmetric polynomial 
$C_\kappa^{(\alpha)}(x_1,\dots, x_N)$ which in turn is proportional to the
Jack polynomial $P_\kappa^{(\alpha)}(x_1,\dots, x_N)$. Thus for
$|x_1|<1, \dots, |x_N|<1$ one defines \cite{Ya92, Ka93}
\begin{equation}\label{4.3a}
{}_2^{} F_1^{(\alpha)}(a,b;c;x_1,\dots,x_N) :=
\sum_\kappa { [a]_\kappa^{(\alpha)} [b]_\kappa^{(\alpha)} \over
[c]_\kappa^{(\alpha)} |\kappa|!} C_\kappa^{(\alpha)}(x_1,\dots, x_N).
\end{equation}  

Let us say a little on the definition of the Jack polynomials
$P_\kappa^{(\alpha)}(x_1,\dots, x_N)=: P_\kappa^{(\alpha)}(x)$ 
and their renormalized version
$C_\kappa^{(\alpha)}(x_1,\dots, x_N)=: C_\kappa^{(\alpha)}(x)$. 
The former are the unique homogeneous
polynomials of degree $|\kappa|$ with the structure
\begin{equation}\label{5.1}
P_\kappa^{(\alpha)}(z) = m_\kappa + \sum_{\mu < \kappa} a_{\kappa \mu}
m_\mu
\end{equation}
(the $a_{\kappa \mu}$ are some coefficients in $\qq(\alpha)$) and which
satisfy the orthogonality
$$
\langle P_\kappa^{(\alpha)}, P_\rho^{(\alpha)} \rangle^{(\alpha)}
\propto \delta_{\kappa, \rho}
$$
where
$$
\langle f, g \rangle^{(\alpha)} :=
\int_{-1/2}^{1/2} d\theta_1 \cdots \int_{-1/2}^{1/2} d\theta_N \,
\overline{f(z_1,\dots,z_N)} g(z_1,\dots,z_N)
\prod_{1 \le j < k \le N} | z_k - z_j |^{2/ \alpha}, \quad
z_j := e^{2 \pi i \theta_j}.
$$
In (\ref{5.1}) $m_\kappa$ denotes the monomial symmetric function (polynomial)
in $x_1,\dots,x_N$ corresponding to $\kappa$, and $\mu < \kappa$ means 
$\sum_{j=1}^p \kappa_j \ge \sum_{j=1}^p \mu_j$ for each $p=1,\dots,N$.
Regarding the definition of $C_\kappa^{(\alpha)}(x_1,\dots,x_N)$, let
\begin{equation}\label{dk}
d_\kappa' = \prod_{(i,j) \in \kappa}
\Big ( \alpha (a(i,j) + 1) + l(i,j)  \Big ),
\end{equation}
where the notation $(i,j) \in \kappa$ refers to the diagram of $\kappa$, in
which each part $\kappa_i$ becomes the nodes $(i,j)$,
$1 \le j \le \kappa_i$ on a square lattice labelled as is conventional
for a matrix. The quantity $a(i,j)$ is the so called arm length (the number of
nodes in row $i$ to the right of column $j$), while $l(i,j)$ is the leg
length (number of nodes in column $j$ below row $i$). In terms of
$d_\kappa'$ we have
\begin{equation}\label{5.1a}
C_\kappa^{(\alpha)}(z) := {\alpha^{|\kappa|} | \kappa|! \over d_\kappa'}
P_\kappa^{(\alpha)}(z).
\end{equation}

\subsection{Integration formulas}
Introduce the Selberg integral
$$
S_N(\lambda_1,\lambda_2,\lambda) := \int_0^1dt_1 \cdots \int_0^1dt_N \,
\prod_{l=1}^N t_l^{\lambda_1}(1 - t_l)^{\lambda_2} \prod_{1 \le j < k \le N}
|t_k - t_j|^{2 \lambda}
$$
and the Morris integral
$$
M_N(a,b,\lambda) := \int_{-1/2}^{1/2}d \theta_1 \cdots  
\int_{-1/2}^{1/2}d \theta_N \, \prod_{l=1}^N e^{\pi i \theta_l (a-b)}
|1 + e^{2 \pi i \theta_l} |^{a+b}
\prod_{1 \le j < k \le N} | e^{2 \pi i \theta_k} - e^{2 \pi i \theta_j}
|^{2 \lambda}
$$
(both these integrals have gamma function evaluations ---
see e.g.~\cite{Fo02} --- although these will not
be needed here). Fundamental to our ability to express (\ref{4.1}) and 
(\ref{1.1a}) in terms of the generalized hypergeometric function 
(\ref{4.3a}) are the generalized Selberg integral evaluation 
\cite{Ma87,Ka97g,Ka93}
\begin{eqnarray}\label{15.64}
&& {1 \over S_N(\lambda_1,\lambda_2,1/\alpha)}
\int_0^1dt_1 \cdots \int_0^1dt_N \,
\prod_{l=1}^N t_l^{\lambda_1}(1 - t_l)^{\lambda_2} 
P_\kappa^{(\alpha)}(t_1,\dots,t_N) 
\prod_{1 \le j < k \le N}|t_k - t_j|^{2/\alpha} \nonumber \\
&& \qquad = P_\kappa^{(\alpha)}(1^N) 
\frac{[\lambda_1 +(N-1)/\alpha +1]^{(\alpha)}_{\kappa} }
{[\lambda_1 + \lambda_2 +2(N-1)/\alpha +2]^{(\alpha)}_{\kappa}}
\end{eqnarray}
and the generalized Morris integral evaluation \cite{Fo95}
\begin{eqnarray}\label{15.63}
&& {1 \over M_N(a,b,1/\alpha)}
\int_{-1/2}^{1/2}d \theta_1 \cdots 
\int_{-1/2}^{1/2}d \theta_N \, \prod_{l=1}^N e^{\pi i \theta_l (a-b)}
|1 + e^{2 \pi i \theta_l} |^{a+b}
P_\kappa^{(\alpha)}(-e^{2 \pi i \theta_1}, \dots, - e^{2 \pi i \theta_N})
\nonumber \\
&& \qquad \times
\prod_{1 \le j < k \le N} | e^{2 \pi i \theta_k} - e^{2 \pi i \theta_j}
|^{2/\alpha} 
= P_\kappa^{(\alpha)}(1^N)\;
\frac{[-b]^{(\alpha)}_{\kappa}}{[1+a+(N-1)/
\alpha]^{(\alpha)}_{\kappa}}
\end{eqnarray}
((\ref{15.64}) and (\ref{15.63}) are in fact equivalent \cite{Ka97g}), where
$$
P_\kappa^{(\alpha)}(1^N) := P_\kappa^{(\alpha)}(x_1,\dots, x_N)\Big 
|_{x_1 = \cdots = x_N = 1}.
$$
Equally important is the generalized binomial summation formula \cite{Ya92,
Ma95}
\begin{equation}\label{15.105}
{}_1^{} F_0^{(\alpha)}(a;x_1,\dots,x_N) :=
\sum_{\kappa} {[a]_\kappa^{(\alpha)} \over |\kappa|!}
C_\kappa^{(\alpha)}(x_1,\dots,x_N) = \prod_{j=1}^N(1 - x_j)^{-a}.
\end{equation}
Thus we see by multiplying both sides of (\ref{15.64}) and  (\ref{15.63})
by
$$
{(x \alpha)^{|\kappa|} [r]_\kappa^{(\alpha)} \over d_\kappa'}
$$
and summing over $\kappa$, using (\ref{15.105}) on the left hand sides
and using the definition (\ref{4.3a}) on the right hand sides together
with the fact that $C_\kappa^{(\alpha)}$ is homogeneous of degree
$|\kappa|$, that \cite{Fo94j}
\begin{eqnarray}\label{15.64a}
&& {1 \over S_N(\lambda_1,\lambda_2,1/\alpha)}
\int_0^1 dt_1 \cdots \int_0^1 dt_N \,
\prod_{l=1}^N t_l^{\lambda_1}(1 - t_l)^{\lambda_2} 
(1 - x t_l)^{-r}
\prod_{1 \le j < k \le N}
|t_k - t_j|^{2/\alpha} \nonumber \\
&& 
\qquad = {}_2^{} F_1^{(\alpha)}(r,
 {1 \over \alpha} (N-1) + \lambda_1 + 1;
{2 \over \alpha} (N-1) + \lambda_1 + \lambda_2 + 2;t_1,\dots,t_N)
 \Big |_{t_1=\cdots=t_N = x}
\end{eqnarray}
and
\begin{eqnarray}\label{15.63a}
&& {1 \over M_N(a,b,1/\alpha)}
\int_{-1/2}^{1/2}d \theta_1 \cdots
\int_{-1/2}^{1/2}d \theta_N \, \prod_{l=1}^N e^{\pi i \theta_l (a-b)}
|1 + e^{2 \pi i \theta_l} |^{a+b}
(1 + x e^{2 \pi i \theta_l} )^{-r}
\nonumber \\
&& \qquad \times 
\prod_{1 \le j < k \le N} | e^{2 \pi i \theta_k} - e^{2 \pi i \theta_j}
|^{2/\alpha} = 
{}_2^{} F_1^{(\alpha)}(r,
-b; {1 \over \alpha} (N-1) +a + 1;
t_1,\dots,t_N) \Big |_{t_1=\cdots=t_N = x},
\end{eqnarray}

\section{Random matrix averages as generalized hypergeometric functions}
\setcounter{equation}{0}
Here we will express (\ref{1.1a}) and (\ref{4.1}) in terms of 
${}_2^{} F_{1}^{(\alpha)}$. Consider first (\ref{1.1a}). By definition
and simple manipulation
\begin{eqnarray*}
\lefteqn{
\Big \langle \prod_{l=1}^N |z - e^{i \theta_l}|^{2 \mu} \Big \rangle_{
{\rm C}\beta {\rm E}_N} }
\nonumber \\
&& \qquad =
{1 \over M_N(0,0,\beta/2) }
\int_{-1/2}^{1/2} d \theta_1 \cdots \int_{-1/2}^{1/2} d \theta_N \,
\prod_{l=1}^N |1 + z e^{2 \pi i \theta_l} |^{2 \mu}
\prod_{1 \le j < k \le N} 
|e^{2 \pi i \theta_k} - e^{2 \pi i \theta_j} |^\beta.
\end{eqnarray*}
Now
$$
| 1 + z e^{2 \pi i \theta_l} |^{2 \mu} = (1 + z e^{2 \pi i \theta_l} )^\mu
(1 + \bar{z} e^{-2 \pi i \theta_l} )^\mu.
$$
Regarding the integrals over $\theta_l \in [-1/2,1/2]$ as parametrizing the unit
circles $w_l = e^{2 \pi i \theta_l}$ in the complex plane, then deforming
each $w_l$ by $w_l \mapsto \bar{z} e^{2 \pi i \theta_l}$ (this is immediate
for $2 \mu, \beta/2 \in \zz_{\ge 0}$ by Cauchy's theorem; it remains valid
for general complex $\mu$ and $\beta$ by Carlson's theorem) we see that we
can write
\begin{eqnarray}\label{7.1}
\lefteqn{
\Big \langle \prod_{l=1}^N |z - e^{i \theta_l}|^{2 \mu} \Big \rangle_{
{\rm C}\beta {\rm E}_N} } 
\nonumber \\ 
&& \qquad =
{1 \over M_N(0,0,\beta/2) }
\int_{-1/2}^{1/2} d \theta_1 \cdots \int_{-1/2}^{1/2} d \theta_N \,
\prod_{l=1}^N |1 + |z|^2 e^{2 \pi i \theta_l} |^{ \mu}
(1 + e^{-2 \pi i \theta_l})^\mu
\nonumber\\
&& \qquad \: \: \times  \prod_{1 \le j < k \le N}
|e^{2 \pi i \theta_k} - e^{2 \pi i \theta_j} |^\beta
\nonumber \\
&& \qquad =
{1 \over M_N(0,0,\beta/2) }
\int_{-1/2}^{1/2} d \theta_1 \cdots \int_{-1/2}^{1/2} d \theta_N \,
\prod_{l=1}^N |1 + |z|^2 e^{2 \pi i \theta_l} |^{ \mu}
e^{- \pi i \theta_l \mu} |1 + e^{-2 \pi i \theta_l}|^\mu \nonumber \\
&& \qquad \: \: \times
\prod_{1 \le j < k \le N}
|e^{2 \pi i \theta_k} - e^{2 \pi i \theta_j} |^\beta
\nonumber \\
&& \qquad =
{}_2^{} F^{(2/\beta)}_1 (-\mu, - \mu; {\beta \over 2} (N-1) + 1;
t_1,\dots, t_N) \Big |_{t_1 = \cdots = t_N = |z|^2}
\end{eqnarray}
where the third equality follows from (\ref{15.63a}), and is valid for
$|z|^2 \le 1$. For $|z|^2 \ge 1$ one notes the simple symmetry
$$
\Big \langle \prod_{l=1}^N |z - e^{i \theta_l}|^{2 \mu} \Big \rangle_{
{\rm C}\beta {\rm E}_N}  = | z|^{2 \mu N}
\Big \langle \prod_{l=1}^N |{1 \over z} - e^{i \theta_l}|^{2 \mu} \Big \rangle_{
{\rm C}\beta {\rm E}_N} ,
$$
thus relating this case back to the case $|z|^2 \le 1$. We remark that
in the case $\beta = 2$ the same generalized hypergeometric function 
evaluation  (\ref{7.1})
has recently been given in \cite{FW03b}. 

Consider next (\ref{4.1}). By definition and simple manipulation, for
$x$ real and greater than or equal to unity
\begin{eqnarray}\label{8.1}
&&\Big \langle \prod_{l=1}^N | x - x_l|^{2 \mu} \Big \rangle_{{\rm J}\beta
{\rm E}_N} \nonumber \\
&& = \qquad
{x^{2 \mu N} \over S_N(a,b,\beta/2)}
\int_0^1 dt_1 \cdots \int_0^1 dt_N \,
\prod_{l=1}^N t_l^a (1 - t_l)^b (1 - t_l/x)^{2 \mu}
\prod_{1 \le j < k \le N} | t_k - t_j|^\beta \nonumber \\
&& = \qquad
x^{2 \mu N} \,
{}_2 F_1^{(2/\beta)}(-2 \mu, {\beta \over 2} (N-1) + a;
\beta (N-1) + a + b + 2; t_1,\dots, t_N)
\Big |_{t_1 = \cdots = t_N = 1/x}
\end{eqnarray}
where the second equality follows from (\ref{15.64a}). The task now is to
analyze the generalized hypergeometric functions in (\ref{7.1}) and
(\ref{8.1}) as $|z| \to 1$ and $x \to 1^+$ respectively. Fortunately, as will
be presented in the next section, the required asymptotic formulas are
available in the literature.

\section{Asymptotic forms}
Using the series definition (\ref{4.3a}), Yan \cite{Ya92} has analyzed the
behaviour of the function ${}_2^{} F_1^{(\alpha)}$ as
$x_1,\dots, x_N \to 1^-$. Thus according to Proposition 4.4 of
 \cite{Ya92} the following result holds, where the notation $A(x) \approx
B(x)$ means that there exists two positive numbers $C_1$ and $C_2$
independent of $x$ such that
$$
C_1 \le {A(x) \over B(x)} \le C_2.
$$

\begin{prop}\label{p1}
Let 
$$
\gamma := a+b-c,
$$
and suppose for all $k$
$$
{ [a]_\kappa^{(\alpha)} [b]_\kappa^{(\alpha)} \over
[c]_\kappa^{(\alpha)} } > 0.
$$
We have for $-1 < x_i < 1$ $(i=1,\dots,N)$

\noindent
(i) If $\gamma > (N-1)\beta/2$, then
\begin{equation}\label{h1}
{}_2^{} F_1^{(\alpha)}(a,b;c;x_1,\dots,x_N) \approx
\prod_{i=1}^N(1 - x_i)^{-\gamma}.
\end{equation} 

\noindent
(ii) If $\gamma < - (N-1)\beta/2$, then there exists a constant $C$ such that
\begin{equation}\label{h2}
{}_2^{} F_1^{(\alpha)}(a,b;c;x_1,\dots,x_N) \le C.
\end{equation}

\noindent
(iii) If $\gamma = \beta(- {N - 1 \over 2} + j -1)$, $j=1,\dots,N$, then for
$x_1=\cdots =x_N=t$,
\begin{equation}\label{h3}
{}_2^{} F_1^{(\alpha)}(a,b;c;x_1,\dots,x_N) \Big |_{x_1=\cdots=x_N=t}
\approx (1-t)^{-(j-1)j \beta/2} \log {1 \over 1 - t}
\end{equation}

\noindent
(iv) If $\beta ( - {N - 1 \over 2} + j - 1 ) < \gamma < \beta
( - {N - 1 \over 2} + j ) $, $j=1,\dots,N-1$, then for
$x_1=\cdots =x_N=t$,
\begin{equation}\label{h4}
{}_2^{} F_1^{(\alpha)}(a,b;c;x_1,\dots,x_N) \Big |_{x_1=\cdots=x_N=t}
\approx (1 -t)^{-j(\gamma + (N-j)\beta/2)}.
\end{equation}
\end{prop}

Application of Proposition \ref{p1} to the ${}_2^{} F_1^{(\alpha)}$ evaluations
(\ref{7.1}) and (\ref{8.1}) gives the desired functional form of the singular
behaviour of (\ref{1.1a}) and (\ref{4.1}) in the case of a negative exponent
$2 \mu$.

\begin{cor} 
Consider the circular ensemble average (\ref{1.1a}).
Suppose $\mu < 0$ and $|z|<1$. For 
$$
2|\mu| = \beta(j-1)+1 \qquad (j=1,\dots,N)
$$
we have
\begin{equation}
\Big \langle \prod_{l=1}^N | z - e^{i \theta_l}
|^{2 \mu} \Big \rangle_{{\rm C}\beta{\rm E}_N} \approx
(1-|z|)^{-(j-1)j\beta/2} \log {1 \over 1 - |z|}
\end{equation}
while for
$$
 \beta(j-1)+1 < 2|\mu| < \beta j + 1 \qquad (j=1,\dots,N)
$$
and thus 
$$
j = {\rm int}[(2|\mu|-1)/\beta + 1], \qquad (2|\mu|-1)/\beta +1 \notin \zz
$$
we have
\begin{equation}\label{e2}
\Big \langle \prod_{l=1}^N | z - e^{i \theta_l}
|^{2 \mu} \Big \rangle_{{\rm C}\beta{\rm E}_N} \approx
(1 - |z|)^{-j(2|\mu|-1-(j-1)\beta/2)}.
\end{equation}

Consider the Jacobi ensemble average (\ref{4.1}).
Suppose $\mu < 0$ and $x > 1$. For 
$$
2|\mu| = \beta (j-1) + 2 + b 
$$
we have
\begin{equation}\label{f1}
x^{-2 \mu N} \Big \langle \prod_{l=1}^N |x - x_l|^{2 \mu} \Big \rangle_{{\rm J}\beta
{\rm E}_N} \approx (1 - 1/x)^{-(j-1)j \beta/2} \log {1 \over 1 - 1/x}
\end{equation}
while for
$$
\beta (j-1) + b + 2 < 2 |\mu| < \beta j + b + 2 \qquad (j=1,\dots,N)
$$
and thus
$$
j = {\rm int}[(2|\mu| - b - 2)/\beta + 1],
\qquad (2|\mu|-b-2)/\beta +1 \notin \zz
$$
we have
\begin{equation}\label{f2}
x^{-2 \mu N} \Big \langle 
\prod_{l=1}^N|x - x_l|^{2 \mu} \Big \rangle_{{\rm J}\beta
{\rm E}_N} \approx (1 - 1/x)^{-j(2|\mu|-b-2-(j-1) \beta / 2)}.
\end{equation}
\end{cor}

The asymptotic behaviour (\ref{e2}) indeed exhibits
the exponent (\ref{2.1}) 
predicted by Berry and Keating \cite{BK02}, and the existence
of a logarithmic correction in the cases that $(2|\mu|-1)/\beta$ is
an integer is confirmed.

Finally, let us apply the results (\ref{f1}) and (\ref{f2}) to deduce the
asymptotic $z \to 1^+$ behaviours of
$$
\Big \langle | \det(zI - U) |^{2 \mu} \Big \rangle_{U \in G}, \quad \mu<0
$$
for $G$ the classical groups $Sp(N)$, $O^+(2N)$ and $O^-(2N)$
(we leave the cases of $O^+(2N-1)$ and $O^-(2N-1)$ as an exercise for the
interested reader). Consider
first $Sp(N)$. The eigenvalues come in complex conjugate pairs
$e^{\pm i \theta_j}$ $(j=1,\dots,N)$ so we have
\begin{equation}\label{12}
\det (zI - U) = (2z)^N \prod_{j=1}^N \Big ( {z^2 + 1 \over 2z} - \cos \theta_j
\Big ).
\end{equation}
Furthermore, we know (see e.g.~\cite{Fo02})
that in the variable $\cos \theta_j =: x_j$ the
eigenvalue probability density function is of the form (\ref{4.2}) with
$a=b=1/2$, $\beta = 2$.
Thus setting $z = 1 + \epsilon$, $0 < \epsilon \ll 1$ 
and noting $(z^2+1)/2z \sim 1 + \epsilon^2/2$ it follows
from (\ref{f1}) that for
$$
2 |\mu| = 2j + 1/2 \qquad (j=1,\dots,N)
$$
we have
\begin{equation}\label{13.1a}
\Big \langle | \det((1 + \epsilon)I  - U) |^{2 \mu} \Big \rangle_{U \in Sp(N)} 
\approx \epsilon^{-2(j-1)j} \log {1 \over \epsilon}
\end{equation}
while for
$$
2j + 1 / 2 < 2 |\mu| < 2(j+1) + 1 / 2 
$$
we have
\begin{equation}\label{13.2}
\Big \langle | \det((1 + \epsilon)I  - U) |^{2 \mu} \Big \rangle_{U \in Sp(N)}
\approx \epsilon^{-2j(2|\mu|-3/2-j)}. 
\end{equation}

For the classical group $O^+(2N)$ the eigenvalues again come in complex
conjugate pairs $e^{\pm i \theta_j}$ $(j=1,\dots,N)$, so (\ref{12})
remains valid. In the variable $\cos \theta_j =: x_j$ the eigenvalue
probability density function is proportional to (\ref{4.2}) with
$a=b=-1/2$, $\beta=2$ (see e.g.~\cite{Fo02}). Thus setting $z=1+\epsilon$, 
$0 < \epsilon \ll 1$ if follows from (\ref{f1}) that for
$$
2|\mu| = 2j -1/2 \qquad (j=1,\dots,N)
$$
we have
\begin{equation}\label{14.1}
\Big \langle | \det((1 + \epsilon)I  - U) 
|^{2 \mu} \Big \rangle_{U \in O^+(2N)}
\approx \epsilon^{-2(j-1)j} \log {1 \over \epsilon}
\end{equation}
while for
$$
2j - 1 / 2 < 2 |\mu| < 2(j+1) - 1 / 2
$$
we have
\begin{equation}\label{14.2}
\Big \langle | \det((1 + \epsilon)I  - U) |^{2 \mu} 
\Big \rangle_{U \in O^+(2N)}
\approx \epsilon^{-2j(2|\mu|-1/2-j)}.
\end{equation}

In the case of the classical group $O^-(2N)$, there is a pair of fixed
eigenvalues at $\pm 1$, with the remaining eigenvalues coming in complex
conjugate pairs $e^{\pm i \theta_j}$ $(j=1,\dots,N-1)$. Hence in this case
\begin{equation}\label{14.3}
\det (zI - U) = (z^2 - 1) (2z)^{N-1} \prod_{j=1}^{N-1} \Big ( {z^2 +1 \over
2z} - \cos \theta_j \Big ).
\end{equation}
In the variable $\cos \theta_j =: x_j$ the eigenvalue probability density
function is proportional to (\ref{4.2}) with $N \mapsto N-1$ and
$a=1/2$, $b=-1/2$, $\beta=2$. As only the value of $b$ enters in (\ref{f1}) and
(\ref{f2}), the sole modification of (\ref{14.1}) and (\ref{14.2})
is multiplication by $\epsilon$ to account for the factor of $z^2-1$ in
(\ref{14.3}). Thus for
$$
2|\mu| = 2j - 1/2 \qquad (j=1,\dots,N-1)
$$
we have
\begin{equation}\label{15.1}
\Big \langle | \det((1 + \epsilon)I  - U) |^{2 \mu} \Big \rangle_{U \in O^-(2N)}
\approx \epsilon^{-2(j-1)j + 1} \log {1 \over \epsilon}
\end{equation}
while for
$$
2j - 1 / 2 < 2 |\mu| < 2(j + 1) - 1/ 2 \qquad (j=1,\dots,N-1)
$$
we have
\begin{equation}\label{15.2}
\Big \langle | \det((1 + \epsilon)I  - U) |^{2 \mu} \Big \rangle_{U \in O^-(2N)}
\approx \epsilon^{-2j(2|\mu|-1/2-j)+1}.
\end{equation}

\section{The macroscopic limit}
\setcounter{equation}{0}
Consider the general Toeplitz determinant
\begin{equation}
D_N[e^{a(\theta)}] := \det \Big [ {1 \over 2 \pi} \int_0^{2 \pi}
e^{a(\theta)} e^{ i (j-k) \theta} \, d \theta \Big ]_{j,k=1,\dots,N}.
\end{equation}
Such determinants are related to averages over $U(N)$ by the simple to
establish formula
\begin{equation}\label{p0}
D_N[e^{a(\theta)}] = \Big \langle \prod_{l=1}^N e^{a(\theta_l)}
\Big \rangle_{U(N)}.
\end{equation}
With $a(\theta)) = \sum_{p=-\infty}^\infty a_p e^{i p \theta}$ and the
Fourier coefficients falling off fast enough that
\begin{equation}\label{p1a}
\sum_{p=-\infty}^\infty |p| |a_p|^2 < \infty,
\end{equation}
an asymptotic formula of Szeg\"o \cite{Sz52} gives
\begin{equation}\label{p2}
D_N[e^{a(\theta)}] \mathop{\sim}\limits_{N \to \infty}
\exp \Big ( Na_0 + \sum_{p=1}^\infty p a_p a_{-p} + {\rm o}(1) \Big ).
\end{equation}
Now, in the case
\begin{equation}\label{p2a}
a(\theta) = 2 \mu \log | z - e^{i \theta} |, \qquad |z| < 1,
\end{equation}
the Fourier coefficients have the explicit form
\begin{equation}\label{p3}
a_p = - {\mu \bar{z}^p \over p}  \: \: (p > 0), \qquad
a_p =  {\mu {z}^{-p} \over p}  \: \: (p < 0), \qquad
a_0=0.
\end{equation}
Since we are assuming $|z| < 1$ these coefficients
satisfy (\ref{p1a}). Substituting
(\ref{p3}) in (\ref{p2}) gives (\ref{1.9}).

In this section, as our final issue, we will make use of a generalization
of (\ref{p2}) due to Johansson to establish the generalization of
(\ref{1.9}) when the average over $U(N)$ is replaced by an average over
C$\beta$E${}_N$. But before doing so, following \cite{HKO01}, we make note
of the interpretation of (\ref{1.9}) as specifying the distribution of
the linear statistic
\begin{equation}\label{q0}
A(z) = \sum_{j=1}^N \log | z - e^{i \theta_j} |^2, \qquad |z| < 1.
\end{equation}
Thus let $P_z(t)$ denote the probability density that $A(z)$ takes on the
value $t$ after averaging over the eigenvalue distribution of $U(N)$. Then
we have
\begin{equation}\label{q0a}
P_z(t) = \Big \langle \delta (t - A(z)) \Big \rangle_{U(N)}
\end{equation}
and consequently
\begin{equation}
\tilde{P}_z(k) := \int_{-\infty}^\infty P_z(t) e^{i t k} \, dt =
\Big \langle \prod_{l=1}^N e^{2 i k \log | z - e^{i \theta_l}|}
\Big \rangle_{U(N)}.
\end{equation}
Thus $\tilde{P}_z(k)$ is precisely (\ref{p0}) with the substitution 
(\ref{p2a}), and $\mu$ therein set equal to $i k$. Consequently
\begin{equation}\label{q1}
\lim_{N \to \infty} P_z(t) = e^{-k^2 \sigma^2/2}, \qquad
\sigma^2 = - \log |1 - |z|^2|^2,
\end{equation}
telling us that the limiting distribution of (\ref{q0}) is a Gaussian
with an $O(1)$ variance taking the explicit value $-\log|1 - |z|^2|^2$.

To study (\ref{q0a}) with the average over $U(N)$ replaced by the probability
density function  C$\beta$E${}_N$, we make note of the following
generalization of the Szeg\"o theorem (\ref{p2}) due to
Johansson \cite{Jo88,Jo98}.

\begin{prop}
Suppose $a(\theta) = \sum_{p=-\infty}^\infty a_p e^{i p \theta}$ and the
Fourier coefficients fall off fast enough that the inequality (\ref{p1a})
holds. We have
\begin{equation}\label{r0}
\Big \langle \prod_{l=1}^N e^{a(\theta_l)}
\Big \rangle_{{\rm C}\beta{\rm E}_N}
\mathop{\sim}\limits_{N \to \infty}
\exp \Big ( Na_0 + {2 \over \beta}
\sum_{p=1}^\infty p a_p a_{-p} + {\rm o}(1) \Big ).
\end{equation}
\end{prop}

\noindent
As an immediate corollary, by substituting (\ref{p3}) for the Fourier
coefficients the sought generalization of (\ref{1.9}) can be deduced.

\begin{cor}
For $|z| < 1$ we have
\begin{equation}\label{r1}
\lim_{N \to \infty}
\Big \langle \prod_{l=1}^N | z - e^{i \theta_l} |^{2 \mu}
\Big \rangle_{{\rm C}\beta{\rm E}_N} =
e^{-(2 \mu^2/ \beta) \log |1 - |z|^2|}.
\end{equation}
\end{cor}

Of course the result (\ref{r1}) affords an interpretation as a fluctuation
formula as in (\ref{q1}), the only difference being that
$\sigma^2 = -(2/\beta) \log |1 - |z|^2|$.

An obvious question is to seek the analogue of (\ref{r1}) for the average
(\ref{4.1}) with $x \notin [0,1]$. Johansson \cite{Jo97} has derived the
analogue of (\ref{r0}), but only in the case $\beta = 2$ and
further with the restriction that $a(\theta)$ is a
polynomial in $\theta$, which is not the case for (\ref{p2a}).

\section*{Acknowledgements}
The work of PJF was supported by the Australian Research Council. We thank
Nalini Joshi for inviting us both to lecture at Sydney University during July
2003 thus facilitating the present collaboration.


\begin{thebibliography}{10}

\bibitem{Be77}
M.V. Berry.
\newblock Focusing and twinkling: critical exponents from catastrophes in
  non-gaussian random short waves.
\newblock {\em J. Phys. A}, 10:2061--2081, 1977.

\bibitem{Be82}
M.V. Berry.
\newblock Universal power-law tails for singularity-dominated strong
  fluctuations.
\newblock {\em J. Phys. A}, 15:2735--2749, 1977.

\bibitem{BK02}
M.V. Berry and J.P. Keating.
\newblock Clusters of near degenerate levels dominate negative moments of
  spectral determinants.
\newblock {\em J. Phys. A}, 35:L1--L6, 2002.

\bibitem{BKS00}
M.V. Berry, J.P. Keating, and H.~Schomerus.
\newblock Universal twinkling exponents for spectral fluctuations associated
  with mixed chaology.
\newblock {\em Proc. R. Soc. A}, 456:1659--1668, 2000.

\bibitem{CF00}
J.B. Conrey and D.W. Farmer.
\newblock Mean values of $L$-functions and symmetry.
\newblock {\em Int. Math. Res. Notices}, 17:883--908, 2000.

\bibitem{CFKRS02}
J.B. Conrey, D.W. Farmer, J.P. Keating, M.O. Rubinstein, and N.C. Snaith.
\newblock Integral moments of zeta- and $L$-functions.
\newblock math.nt/0206018.

\bibitem{Fo02}
P.J. Forrester.
\newblock Log-gases and {Random} {Matrices}.
\newblock www.ms.unimelb.edu.au/\~{}matpjf/matpjf.html.

\bibitem{Fo94j}
P.J. Forrester.
\newblock Addendum to {Selberg} correlation integrals and the $1/r^2$ quantum
  many body system.
\newblock {\em Nucl. Phys. B}, 416:377--385, 1994.

\bibitem{Fo95}
P.J. Forrester.
\newblock Integration formulas and exact calculations in the
  {Calogero-Sutherland} model.
\newblock {\em Mod. Phys. Lett B}, 9:359--371, 1995.

\bibitem{FW03b}
P.J. Forrester and N.S. Witte.
\newblock Discrete {P}ainlev\'e equations, orthogonal polynomials on the unit
  circle and $n$-recurrences for averages over {$U(N)$} -- {$P_{\rm VI}$}
  $\tau$-functions.
\newblock math-ph/0308036.

\bibitem{FK03}
Y.V. Fyodorov and J.P. Keating.
\newblock Negative moments of characteristic polynomials of random {GOE}
  matrices and singularity dominated strong fluctuations.
\newblock {\em J. Phys. A}, 36:4035--4046, 2003.

\bibitem{HKO01}
C.P. Hughes, J.P. Keating, and N.~O'Connell.
\newblock On the characteristic polynomial of a random unitary matrix.
\newblock {\em Commun. Math. Phys.}, 220:429--451, 2001.

\bibitem{Jo88}
K.~Johansson.
\newblock On Szeg\"o's formula for {Toeplitz} determinants and generalizations.
\newblock {\em Bull. Sc. Math., $2^{\rm e}$ s\'erie}, 112:257--304, 1988.

\bibitem{Jo97}
K.~Johansson.
\newblock On random matrices from the compact classical groups.
\newblock {\em Ann. of Math.}, 145:519--545, 1997.

\bibitem{Jo98}
K.~Johansson.
\newblock On fluctuation of eigenvalues of random {Hermitian} matrices.
\newblock {\em Duke Math. J.}, 91:151--204, 1998.

\bibitem{Ka97g}
K.W.J. Kadell.
\newblock The {Selberg-Jack} symmetric functions.
\newblock {\em Adv. Math.}, 130:33--102, 1997.

\bibitem{Ka93}
J.~Kaneko.
\newblock Selberg integrals and hypergeometric functions associated with {Jack}
  polynomials.
\newblock {\em SIAM J. Math Anal.}, 24:1086--1110, 1993.

\bibitem{KP01}
J.P. Keating and S.D. Prado.
\newblock Orbit bifurcations and the scarring of wave functions.
\newblock {\em Proc. R. Soc. A}, 457:1855--1872, 2001.

\bibitem{KS00b}
J.P. Keating and N.C. Snaith.
\newblock Random matrix theory and {$L$}-functions at $s=1/2$.
\newblock {\em Commun. Math. Phys.}, 214:91--110, 2000.

\bibitem{KS00a}
J.P. Keating and N.C. Snaith.
\newblock Random matrix theory and $\zeta(1/2 + it)$.
\newblock {\em Commun. Math. Phys.}, 214:57--89, 2001.

\bibitem{KS03}
J.P. Keating and N.C. Snaith.
\newblock Random matrices and {$L$}-functions.
\newblock {\em J. Phys. A}, 36:2859--2881, 2003.

\bibitem{Ma87}
I.G. Macdonald.
\newblock Commuting differential operators and zonal spherical functions.
\newblock In A.M.~Cohen et~al., editor, {\em Algebraic Groups, Utrecht 1986},
  volume 1271 of {\em Lecture Notes in Math.}, pages 189--200. Springer-Verlag,
  Heidelberg, 1987.

\bibitem{Ma95}
I.G. Macdonald.
\newblock {\em Hall polynomials and symmetric functions}.
\newblock Oxford University Press, Oxford, 2nd edition, 1995.

\bibitem{Sz52}
G.~Szeg\"o.
\newblock On certain {H}ermitian forms associated with the {F}ourier series of
  a positive function.
\newblock {\em Comm. Seminaire Math. de l'Univ. de Lund}, 
\newblock tome suppl\'ementaire, d\'edi\'e \`a Marcel Riesz,
pp. 228--237, 1952.

\bibitem{Ya92}
Z.~Yan.
\newblock A class of generalized hypergeometric functions in several variables.
\newblock {\em Can J. Math.}, 44:1317--1338, 1992.

\end{thebibliography}

\end{document}